\documentclass[12pt]{article}

\usepackage{amsmath,amsfonts,latexsym,amssymb,amscd}
\usepackage{pslatex}
\usepackage[latin1]{inputenc}
\usepackage[T1]{fontenc}
\usepackage{pspicture}
\usepackage{verbatim,amsthm,curves,graphics}
\usepackage{mathrsfs}

\usepackage{graphicx}
\renewcommand{\v}{{\bf v}}
\newcommand{\G}{{\mathcal G}}
\newcommand{\mr}{{\mathbb R}}

\newcommand{\mc}{{\mathbb C}}
\newcommand{\mz}{{\mathbb Z}}

\newcommand{\K}{{\mathcal K}}

\begin{document}


\title{A Uniqueness theorem for 5-dimensional Einstein-Maxwell black holes}

\author{
Stefan Hollands$^{1}$\thanks{\tt HollandsS@Cardiff.ac.uk}\:,
Stoytcho Yazadjiev$^{2,3}$\thanks{\tt yazadj@theorie.physik.uni-goe.de}\:,
\\ \\
{\it ${}^{1}$ School of Mathematics, Cardiff University} \\
{\it Cardiff, CF24 4AG, UK} \medskip \\
{\it ${}^{2}$Institut f\"ur Theoretische Physik,
     Universit\"at G\" ottingen,} \\
{\it D-37077 G\" ottingen, Germany} \medskip \\
{\it ${}^{3}$Department of Theoretical Physics, Faculty of Physics, Sofia
University} \\
{\it 5 J. Bourchier Blvd., Sofia 1164, Bulgaria} \\
    }
\date{}

\maketitle

\begin{abstract}
In a previous paper [gr-qc/0707.2775] we showed that stationary
asymptotically flat vacuum black hole solutions in 5 dimensions
with two commuting axial Killing fields can be completely characterized by
their mass, angular momentum, a set of real moduli, and a set
of winding numbers. In this paper we generalize our analysis to
include Maxwell fields.
\end{abstract}


\sloppy

\section{Introduction}

In $n=4$ spacetime dimensions, asymptotically flat, stationary vacuum or electrovac
black hole solutions are completely characterized by their asymptotic charges---mass, angular
momentum, and electric charge~\cite{Carter,Robinson,Mazur,Bunting}.
The complete classification of stationary black holes in more than $n=4$ spacetime dimensions
is at present an open problem. However, in a recent paper of ours~\cite{us}, a partial
classification was achieved for vacuum solutions under the assumption that the number of
commuting axial\footnote{By this we mean a Killing field whose orbits are periodic.} Killing fields
is sufficiently large. The particular case considered there was
$n=5$, and the number of axial Killing fields required was two\footnote{The higher dimensional
rigidity theorem~\cite{HIW} only gives one extra axial Killing field. This is presumably the
generic situation.}.
Under this hypothesis, we showed how to construct from the given solution a certain set of invariants
consisting of a set of real numbers ("moduli") and a collection of integer-valued vectors
("winding numbers"). These data were called the "interval structure" of the solution. It
determines in particular the horizon topology, which could be either
$S^3, S^1 \times S^2$ or a Lens-space $L(p,q)$. We then demonstrated that the interval structure together
with the asymptotic charges gives a complete set of invariants of the solutions, i.e.,
if these data coincide for two given solutions, then the solutions are isometric.

In this paper, we generalize the analysis of our previous paper~\cite{us} to include Maxwell fields.
We show that, if certain restrictive additional conditions are imposed upon the Maxwell field
and the axial Killing fields, then a similar uniqueness theorem as in the vacuum case can be
proven. Namely, we find that the solution is now completely characterized by the interval structure,
the magnetic charges, as well as the mass and angular momentum. The extra assumptions placed upon
the Killing fields imply that the electric charge (but not the magnetic charges), and
one of the angular momenta vanishes. They also imply that the possible interval structures are
limited. In particular, the horizon topology can only be either $S^3$ or $S^2 \times S^1$,
but not $L(p,q)$.

Non-trivial Einstein-Maxwell black rings (horizon $S^1 \times S^2$) satisfying our assumptions\footnote{
Note that the Einstein-Maxwell black ring found in~\cite{Elvang} has non-vanishing 
electric charge and hence does not fall into the class studied in the present paper.} have been
found by~\cite{REmparan} (see also~\cite{Yazadjiev} ).

\section{Stationary Einstein-Maxwell black holes in $n$ dimensions}

Let $(M,g_{ab},F_{ab})$ be an $n$-dimensional, analytic, asymptotically flat,
stationary black hole spacetime satisfying the Einstein-Maxwell
equations
\begin{eqnarray}
&&R_{ab}= \frac{1}{2}\left(F_{ac}F_{b}{}^c- \frac{g_{ab}}{2(n-2)}F_{cd}F^{cd} \right),\\
&&\nabla_{a}F^{ab}=0=\nabla_{[a} F_{bc]}.
\end{eqnarray}
Let $t^a$ be the
asymptotically timelike Killing field, $\pounds_t g_{ab} = 0$,
which we assume is normalized
so that $\lim \,  g_{ab} t^a t^b = -1$ near infinity. We assume that also
the Maxwell tensor is invariant under $t^{a}$, in the sense that $\pounds_t F_{ab} = 0$.
We denote by
$H = \partial B$ the horizon of the black hole, where the black hole $B$
is defined as usual by $B = M \setminus
I^-({\mathcal J}^+)$, with ${\mathcal J}^\pm$ the null-infinities of
the spacetime. It is assumed that the latter has topology $\mr \times \Sigma_\infty$, where
$\Sigma_\infty$ is metrically and topologically an $(n-2)$-dimensional sphere.\footnote{In 4
  dimensions, $\Sigma_\infty$ may be {\em shown} to be an $S^2$ under
  suitably strong additional hypothesis. A discussion of the structure
of null-infinity in higher dimensions is given in~\cite{HI}.}
We assume that $H$ is ``non-degenerate''
and that the horizon cross section is a compact connected
manifold of dimension $n-2$. Under these conditions, one
of the following 2 statements is true: (i) If $t^a$ is tangent to the
null generators of $H$ then the spacetime must be static~\cite{Sudarsky}.
(ii) If $t^a$ is not tangent to the
null generators of $H$, then the higher dimensional rigidity
theorem~\cite{HIW} states that there
exist $N\ge1$  additional linear independent, mutually commuting Killing fields
$\psi_{1}^a, \dots, \psi^a_{N}$, such that
$\pounds_{\psi_{1}} F_{ab}\dots, \pounds_{\psi_{N}} F_{ab} = 0$.
These Killing fields generate
periodic, commuting flows (with period $2\pi$), and there exists a
linear combination
\begin{equation}
K^a = t^a + \Omega_{1}^{} \psi_{1}^a + \dots +
\Omega_{N}^{} \psi_{N}^a, \quad \Omega^{}_{i} \in \mr
\end{equation}
so that the Killing field $K^a$ is tangent and normal to the null
generators of the horizon $H$, and
\begin{equation}\label{orth}
K_a \psi_{i}^a = 0 \quad \text{on $H$.}
\end{equation}
Thus, in case (ii), the spacetime
is axisymmetric, with isometry group $\G=\mr \times U(1)^N$.
From $K^a$, one may define the surface gravity of the black hole by
$\kappa^2 = \lim_H (\nabla_a f) \nabla^a f/f$, with $f=(\nabla^a K^b)
\nabla_a K_b$ the norm, and it may be shown that $\kappa$ is constant
on $H$~\cite{Waldbook}. In fact, the non-degeneracy condition implies
$\kappa > 0$.

In case (i), one can prove that the spacetime is actually unique,
and in fact isometric to the Reissner-Nordstr\" om-Tangherlini
spacetime~\cite{Israel67}, for higher dimensions see~\cite{Gibbons}.
In this paper, we will be concerned with case (ii).

Similar to 4 dimensions, the mass and angular momenta
of the solution associated with the Killing fields are given, up to irrelevant
numerical factors, by the Komar expressions
\begin{equation}
m= -\frac{n-2}{n-3} \int_{\Sigma_\infty} \nabla_a t_b \, dS^{ab} \, , \quad
J_i = \int_{\Sigma_\infty} \nabla_a \psi_{i \, b} \, dS^{ab}
\end{equation}
and we define the electric and magnetic charges of the solution by
\begin{equation}
Q_{\rm E}[\Sigma_\infty] = \int_{\Sigma_\infty} F_{ab} \, dS^{ab} \, ,\quad
Q_{\rm M}[C_l] = \int_{C_l} *F_{ab \dots c} \, dS^{ab \dots c}
\, ,
\end{equation}
where $C_l, l=1,2,\dots$ runs through all the topologically inequivalent, non-contractible, closed
2-surfaces in the exterior of the spacetime.
These numbers are invariants of the solution, and in 4 dimensions in fact
characterize the solution uniquely. However, in higher dimensions this is no longer the
case. In fact, we will see that further invariants must be taken into account as well.

We now restrict attention to the exterior of the black hole,
$I^-({\mathcal J}^+)$, which we shall again denote by $M$ for
simplicity. We assume that the exterior $M$ is globally hyperbolic.
By the topological censorship theorem~\cite{Woolgar},
the exterior $M$ is a simply connected manifold (with boundary
$\partial M = H$). To understand better the nature of the
solutions, it is useful to first eliminate the coordinates
corresponding to the symmetries of the spacetime.
More precisely, one considers the factor space $\hat M = M/\G$, where
$\G$ is the isometry group of the spacetime generated by the Killing
fields. Since
the Killing fields $\psi^a_{i}$ in general have zeros, the
factor space $\hat M = M/\G$ will normally have singularities and is
difficult to analyze. However, when the number of axial Killing fields is
equal to $N=n-3$, and if there are no points in the exterior $M$ whose isotropy
subgroup is discrete, then the factor space can be analyzed by elementary means.
This analysis was carried out in~\cite{us} for the case of $n=5$, and a very similar
analysis also applies to general $n$. Since we are assuming that
the spacetime is asymptotically flat in the
standard sense with spherical infinity $\Sigma_\infty \cong S^{n-2}$,
the group of asymptotic symmetries with compact orbits
must be isomorphic to a subgroup of $SO(n-1)$, whose maximal torus has dimension
$[(n-1)/2]$. Thus $n-3$ axial Killing fields are
only possible if either $n=4$, or if $n=5$. From now on, we focus on the latter case.

Thus, from now on we assume that the
isometry group of the spacetime is ${\cal G}={\cal K}\times \mr$, where ${\cal K}=U(1)\times U(1)$, and we also assume that the action of the isometry group ${\cal K}$ generated by the axial symmetries is so that there are no points with discrete isotropy group. We denote the Killing vector fields generating $\cal K$ by $\psi_1^a, \psi^a_2$, and
we denote the factor space $\hat M = M/{\cal G}$.  The nature of the factor space is described by the following
proposition~\cite{us}:

\medskip
\noindent
{\bf Proposition 1:} Let $(M,g_{ab})$ be the exterior of
a stationary, asymptotically flat, Einstein-Maxwell black hole spacetime
with $2$ mutually commuting independent axial
Killing fields $\psi^a_{1}, \psi_2^a$. Then the
orbit space $\hat M=M/\G$ by the isometry group
is a simply connected, 2-dimensional manifold with boundaries and
corners. Points in the interior of $\hat M$ correspond to points in $M$ where
all Killing fields $t^a, \psi_1^a, \psi_2^a$ are
linearly independent. Points on the $i$-th 1-dimensional boundary segment of $\partial \hat M$
correspond to either the horizon of $M$, or points where a linear combination
$v^1_i \psi_1^a + v^2_i \psi^a_2 = 0$, where ${\bf v}_i = (v^1_i, v^2_i)$ is
a vector of integers that is constant on each such segment. Points in
the corners of $\partial \hat M$ correspond to points in $M$ where $\psi_1^a = 0 = \psi_2^a$.
The boundary of $\hat M$ is connected.

\medskip
\noindent
Away from the boundary of $\hat M$, we can define a metric $\hat g_{ab}$ by identifying
the tangent space $T_{\pi (x)} \hat M$ with the subspace $H_x$ of $T_x M$ spanned by the
vectors orthogonal to $t^a, \psi^a_1, \psi^a_2$, where $\pi: M \to \hat M=M/{\cal G}$ is the
projection. We denote this metric by $\hat g_{ab}$. It has signature $(++)$. We denote
the derivative operator associated with this metric by $\hat D_a$. If one defines the $3\times 3$
Gram matrix of the Killing fields by
\begin{equation}
G_{IJ} = g_{ab} X^a_I X^b_J, \quad
X_I^a =
\begin{cases}
t^a_{} & \text{if $I=0$}, \\
\psi_{i}^a & \text{if $I=1,2$,}
\end{cases}
\end{equation}
then the Gram determinant
\begin{equation}
r^2 = |{\rm det} \, G|
\end{equation}
defines a scalar function $r$ on $\hat M$ which is harmonic,
$\hat D^a \hat D_a r = 0$, as a consequence of the Einstein-Maxwell
equations. Using this, one can show that $r > 0, \hat D_a r \neq 0$ on the interior of $\hat M$, and
one can also show that $r=0$ on $\partial \hat M$. A conjugate harmonic scalar field $z$ may
then be defined on $\hat M$ by the equation $\hat D_a z = \hat \epsilon^b{}_a \hat D_b r$.
The functions $r,z$ define global coordinates on $\hat M$, thus identifying this space with
the complex upper half-plane
$$\hat M = \{ \zeta = z + ir \in \mc: \,\,\, r \ge 0\} \, ,$$
with the boundary segments corresponding to intervals on the real axis. The length $z_i - z_{i+1} = l_i$ of each
segment is an invariant of the solution. The induced metric $\hat g_{ab}$ is given in these coordinates by
\begin{equation}
d\hat s^2 = k(r,z)^2 (dr^2 + dz^2)
\end{equation}
with $k^2$ a conformal factor.

The set of real "moduli" $\{l_i\}$, and and of the "winding number" vectors $\{{\bf v}_i\}$
are global parameters that can be defined in an invariant way for the given solution in addition to
the mass $m$, the two angular momenta $J_1, J_2$, and the electric and magnetic charges.
We refer to these data as the ``interval structure'' of the solution. As shown in~\cite{us},
the interval structure determines the structure of $M$ as a fibered space with an action of
the torus group $\K$. The winding numbers $\{{\bf v}_i\}$ characterize the structure of this
fibration near the axis segments. It follows from our analysis in~\cite{us} that near such an axis,
$M$ locally has the structure of $\mr^2 \times {\rm Seiffert}(v^1_i, v^2_i)$, i.e., it is a cartesian
product of $\mr^2$ with a Seiffert torus, i.e., a 3-torus with a twisting characterized by the two
winding numbers. The winding numbers on segments adjacent on a corner, respectively adjacent
on the horizon have to satisfy the constraint~\cite{us}

\begin{center}\label{table}
\begin{tabular}{l|c}
${\rm det} \, (\v_j, \v_{j+1}) = \pm 1$ & if $(z_{j-1},z_j)$
and $(z_j, z_{j+1})$ are not the horizon \\
\hline
${\rm det} \, (\v_{h-1}, \v_{h+1}) = p$ & if $(z_h,
  z_{h+1})$ is the horizon \\
\end{tabular}
\end{center}

Furthermore, we have the following theorem about the horizon topology~\cite{us}:

\medskip
\noindent
{\bf Proposition 2:} In a black hole spacetime of dimension $5$
with $2$ commuting, independent axial Killing fields, the horizon cross section $\mathcal H$
must be topologically either a ring $S^1 \times S^2$, a sphere $S^3$, or
a Lens-space $L(p,q)$, with $p,q \in \mz$, and $p$ as in eq.~\eqref{table}.

\medskip
\noindent
{\bf Remark:} The Lens-spaces $L(p,q)$
(see e.g.~\cite[Paragraph 9.2]{Adams}) are the
spaces obtained by gluing the boundaries of two solid tori together
in such a way that the meridian of the first goes to a curve on the
second which wraps around the longitude $p$-times and which wraps
around the meridian $q$-times. A Lens-space may also be obtained as
the quotient of $S^3$ by a discrete group of isometries.

\medskip
\noindent
For illustrative purposes, we list the interval structure for some known solutions~\cite{Myers,Emparan,Elvang,Pom}:

\vspace{1cm}
\begin{center}
\begin{tabular}{|c|c|c|c|}
\hline
               & Moduli $l_i$ & Vectors ${\bf v}_i$ & Horizon Topology \\
\hline
Myers-Perry BH & $\infty, l_1, \infty $ & $(1,0),
(0,0), (0,1)$ & $S^3$ \\
\hline
Black Ring & $ \infty, l_1, l_2, \infty $ & $(1,0),
(0,0), (1,0), (0,1)$ & $S^2 \times S^1$\\
\hline
Flat Spacetime & $\infty, \infty $ & $(1,0), (0,1)$ & ---\\
\hline
\end{tabular}
\end{center}

\noindent
Here we are using the convention that the integer vector ${\bf v}_h$ associated
with the horizon is taken to be $(0,0)$. Even for
a fixed set of of asymptotic charges $m,J_1,J_2$ the invariant lengths
$l_1,l_2$
may be different for the different Black Ring solutions,
corresponding to the fact that there exist non-isometric
Black Ring solutions with equal asymptotic charges.

\newpage

\section{Moduli space of Einstein-Maxwell black holes}

We would now like to see to what extent the interval structure, and
the global charges $m, J_1, J_2, Q_{\rm E}, Q_{\rm M}$ determine a given black hole solution
of the Einstein-Maxwell equations in 5 dimensions. We were not able to
analyze this question in generality but only in a simplified case. The simplifying
assumptions that we will make in this section in addition to the general hypothesis
stated above are the following:
\begin{enumerate}
\item About the {\em spacetime metric} we assume that one of the axial Killing fields, say $\psi_1^a$, is orthogonal to the other Killing fields, $g_{ab} \psi_1^a \psi_2^b = 0 = g_{ab} t^a \psi_1^b$,
    and that it is hypersurface orthogonal, $\psi_{1 [a} \nabla_b \psi_{1 c]} = 0$.
\item About the {\em Maxwell field} we assume that there is a 1-form $\xi_a$ orthogonal
      to the Killing fields such that $F_{ab} = \xi_{[a} \psi_{1 b]}$.
      It can easily be shown that, if the Maxwell field arises from a vector potential
      $F_{ab} = 2\nabla_{[a} A_{b]}$ which is invariant under the Killing fields, then this will be the case if and only if $A^a$ is proportional to $\psi_1^a$ at each point in $M$.
      Note, however that we do {\em not} assume the existence of such a vector potential here.
\end{enumerate}

Let us first point out some simplifications which follow from assumptions
1) and 2). The first immediate consequence of 1) is that $J_1=0$.
Secondly, because the Killing field $\psi_1^a$ is demanded to be orthogonal to $\psi_2^a$,
if $v^1 \psi_1^a + v^2 \psi_2^a = 0$ at a point in spacetime, then either ${\bf v} = (v^1, v^2) = (0,0)$,
or ${\bf v} = (0,1), (1,0)$, or both axial Killing fields vanish. Thus, the interval structure (see Prop.~1) of
any solution satisfying assumption 1) can only be of the following possibilities (i)---(iv):

\vspace{1cm}
\begin{center}
\begin{tabular}{|c|c|c|}
\hline
               & Moduli $l_i$ & Vectors ${\bf v}_i$ \\
\hline
(i) & $\infty, l_1, \dots, l_p, \infty $ & $(1,0), (0,1), \dots (1,0), (0,0), (1,0), (0,1)\dots, (0,1)$ \\
\hline
(ii) & $\infty, l_1, \dots, l_p, \infty $ & $(1,0), (0,1), \dots (0,1), (0,0), (0,1), (1,0) \dots, (0,1)$ \\
\hline
(iii) & $\infty, l_1, \dots, l_p, \infty $ & $(1,0), (0,1), \dots (0,1), (0,0), (1,0), (0,1) \dots, (0,1)$ \\
\hline
(iv) & $\infty, l_1, \dots, l_p, \infty $ & $(1,0), (0,1), \dots (1,0), (0,0), (0,1), (1,0) \dots, (0,1)$ \\
\hline
\end{tabular}
\end{center}

Thus, the possible interval structures are severely restricted by 1). By Prop.~2, it then follows
that the only possible horizon topologies are
\begin{equation}
{\mathcal H} \cong S^1 \times S^2 \quad \text{(black ring)}, \quad {\mathcal H} \cong S^3 \quad \text{(black hole),}
\end{equation}
with the first case realized when the vectors to the left and right of the horizon
${\bf v}_{h-1}, {\bf v}_{h+1}$ are equal [i.e., for the interval structures (i) and (ii)] and the
second case realized when they are different [i.e., for the interval structures (iii) and (iv)].
In particular, the Lens-spaces $L(p,q)$ are excluded as possible horizon topologies by 1).

From 2), the electric charge vanishes, $Q_{\rm E} = 0$, and the Maxwell field is completely characterized by the 1-form
\begin{equation}
f_a =  F_{ab} \psi^a_1 \, ,
\end{equation}
which is closed by the equations of motion for the Maxwell field,
$\nabla_{[a} f_{b]} = 0$. We define the twist 1-form by
\begin{eqnarray}
\omega_a ={1\over 2} \varepsilon_{abcde}\psi^{b}_1 \psi^{c}_2 \nabla^{d}\psi_2^e \, .
\end{eqnarray}
Using that $\psi_1^a$ and $\psi_2^a$ are commuting Killing fields,
we find that $\nabla_{[a} \omega_{b]}$ is proportional to
$\epsilon_{abcde} \psi_2^c \psi_1^{e} R^{df} \psi_{2 \, f}$.
If we now substitute the Einstein-Maxwell equation for the Ricci tensor, and use
assumptions 1) and 2), then we see that $\nabla_{[a} \omega_{b]} = 0$.
By definition, $\omega_a$ and $f_a$ are invariant under the symmetries,
so they induce corresponding 1-forms $\hat \omega_a$ and $\hat f_a$ on the factor space $\hat M$, which are
still closed. Since the factor space is the upper half plane $\{ \zeta = z+ir: \,\,\, r \ge 0\}$,
i.e. is in particular simply connected, we can define
global potentials for these quantities, $\hat D_a \chi = \hat \omega_a$ and $\hat D_a \alpha = \hat f_a$.
If the Maxwell field arises from a globally defined vector potential, $F_{ab}=2\nabla_{[a} A_{b]}$---which we do {\em not} assume---then $\alpha = A_a \psi_1^a$.

\medskip
\noindent
Using the potentials $\alpha, \chi$, we can now write down the reduced Einstein-Maxwell equations on the orbit space $\hat M$.
Let $\nu, w, u$ be the functions on $\hat M$ be defined by
\begin{equation}
e^{2u} = g_{ab} \psi_1^a \psi_1^b \, , \quad e^{-u+2w} = g_{ab} \psi_2^a \psi_2^b \, , \quad
e^{-u+2w+2\nu} = (\nabla_a r) \nabla^a r \, .
\end{equation}
Then the complete Einstein-Maxwell equations are equivalent to the following set of
equations on the upper complex half plane $\hat M$~\cite{Yazadjiev} :
\begin{eqnarray}\label{smodel}
\hat D^a \left(r \Phi^{-1}_{1} \hat D_{a} \Phi_{1}^{}\right)&=&0 \, ,\nonumber\\
\hat D^a \left(r \Phi^{-1}_{2} \hat D_{a} \Phi_{2}^{}\right)&=&0 \, ,
\end{eqnarray}
together with
\begin{eqnarray}\label{nudef}
-r^{-1} (\hat D^a r) \hat D_a \nu &=&
\left[
{3\over 8} {\rm Tr} \left(\hat D^a \Phi_{1}^{} \hat D^b \Phi^{-1}_{1}\right)+
{1\over 8} {\rm Tr} \left(\hat D^a \Phi_{2}^{} \hat D^b \Phi^{-1}_{2}\right)
\right] \cdot
\left[ \hat g_{ab} - 2(\hat D_a z) \hat D_b z \right] \nonumber \\
-r^{-1} (\hat D^a z) \hat D_a \nu &=&
\left[
{3\over 4} {\rm Tr}\left( \hat D^a \Phi_{1}^{} \hat D^b \Phi^{-1}_{1} \right)+
{1\over 4} {\rm Tr}\left( \hat D^a \Phi_{2}^{} \hat D^b \Phi^{-1}_{2} \right)
\right] (\hat D_a r) \hat D_b z \, ,
\end{eqnarray}
where the matrix fields are defined in terms of $u, w,\alpha,\chi$ by
\begin{eqnarray}
\Phi_{1} = \left(%
\begin{array}{cc}
  e^{u} + {1\over 3}e^{-u}\alpha^2 & {1\over \sqrt{3}} e^{-u}\alpha \\
 {1\over \sqrt{3}} e^{-u}\alpha & e^{-u} \\\end{array}%
\right) ,
\end{eqnarray}
and
\begin{eqnarray}
\Phi_{2} = \left(%
\begin{array}{cc}
  e^{2w} + 4\chi^2e^{-2w} & 2\chi e^{-2w} \\
 2\chi e^{-2w} & e^{-2w} \\\end{array}%
\right).
\end{eqnarray}
The first two equations state that
the matrix fields $\Phi_{1}$ and $\Phi_{2}$ each satisfy the equations of a 2-dimensional
sigma-model. The matrix fields are real, symmetric, with determinant equal to $1$ on the interior of $\hat M$.
We may view them as taking values in the hyperbolic space $\mathbb H$. The matrix fields $\Phi_1, \Phi_2$
determine the functions $\alpha, \chi, w, u$.
The second and third equations~\eqref{nudef} are decoupled from the
sigma-model equations and determine the function $\nu$.

Using this formulation of the reduced Einstein-Maxwell equations, we will now prove the
main result of this paper:

\medskip
\noindent
{\bf Theorem:} Consider two stationary, asymptotically flat, Einstein-Maxwell
black hole spacetime of dimension 5,  having  one
time-translation Killing field and two axial Killing fields.
We also assume that there are no points with discrete isotropy subgroup under the action of the
isometry group in the exterior of the black hole, and we
assume that the Killing and
Maxwell fields satisfy the assumptions 1) and 2) above, implying that
${\bf v}_i = (1,0)$ or $(0,1)$, and ${\mathcal H} = S^3$ or $S^1 \times S^2$, and $Q_{\rm E} = 0 = J_1$ for the solutions. If
the two solutions have the same interval structures, the same
values of the mass $m$, same angular momentum  $J_2$, and same
magnetic charges $Q_{\rm M}[C_l]$ for all 2-cycles $C_l$, then they are isometric.

\medskip
\noindent
{\em Proof:}
Consider two solutions $(M,g_{ab},F_{ab})$ and $(\tilde M, \tilde g_{ab}, \tilde F_{ab})$
as in the statement of the
theorem. As argued in~\cite{us}, since the interval structures of both solutions are the same,
$M$ and $\tilde M$ can be identified as manifolds, and the actions of the isometry group $\mathcal G$
are conjugate to each other. Thus, we may assume that $\tilde M = M$, and that $\tilde t^a = t^a$,
$\tilde \psi^a_i = \psi^a_i$. Furthermore, since the quotient space by the isometries is the upper half plane in both cases, we may assume that $\tilde r = r, \tilde z = z$ as functions on $\tilde M = M$. We now define the
2 by 2 matrix fields as above, which we denote $\tilde \Phi_i$ and $\Phi_i$, $i=1,2$. These functions are
mappings $\hat M \to {\mathbb H}$ from the upper complex half plane into the 2-dimensional hyperbolic space.
We next consider the functions
\begin{equation}\label{sigdef1}
\sigma_1 = {\rm Tr} \Big[ \Phi_1^{-1} \tilde \Phi_1^{} - 1 \Big] =  \frac{(e^{u}-e^{\tilde u})^2}{e^{u}e^{\tilde u}} + \frac{1}{3}
\frac{\left(\tilde \alpha - \alpha \right)^2}{e^{u}e^{\tilde u}}
\end{equation}
and
\begin{equation}\label{sigdef2}
\sigma_2 = {\rm Tr} \Big[ \Phi_2^{-1} \tilde \Phi_2^{} - 1 \Big] =  \frac{(e^{2w}-e^{2\tilde w})^2}{e^{2w}e^{2\tilde w}} + 4
\frac{\left(\tilde \chi - \chi \right)^2}{e^{2w}e^{2\tilde w}} \, .
\end{equation}
The quantity $\sigma_1$ is a function of the point wise
geodesic distance between the maps $\Phi_1$ and $\tilde \Phi_1$ in the target space $\mathbb H$, and $\sigma_2$ similarly
between $\Phi_2$ and $\tilde \Phi_2$.
By a straightforward calculation using the equations~\eqref{smodel}, one finds that
the functions $\sigma_i$ satisfy the differential inequality
\begin{equation}\label{subharm0}
\hat D^a (r \hat D_a \sigma_i) \ge 0 \, , \quad \text{for $i=1,2$.}
\end{equation}
It is now convenient to view the maps $\sigma_i$ not as functions on the complex upper half plane
$\hat M = \{ \zeta = z + ir \in \mc: \,\,\, r \ge 0 \}$,
but as axially symmetric functions on
$\mr^3 \setminus \{ z-{\rm axis} \}$, by writing points $X=(X_1, X_2, X_3) \in \mr^3$
in cylindrical coordinates as $X = (r \cos \varphi, r \sin \varphi, z)$. Eqs.~\eqref{subharm0}
may then be written as
\begin{equation}\label{subharm}
\left\{
\frac{\partial^2}{\partial X_1^2} +
\frac{\partial^2}{\partial X_2^2} +
\frac{\partial^2}{\partial X_3^2}
\right\} \sigma_i(X) \ge 0 \, , \quad \text{for $i=1,2$.}
\end{equation}
By a general arguments based on the maximum principle, see e.g.~\cite{Weinst1,Weinst2},
if $\sigma_i$ are globally bounded above on the
entire $\mr^3$ including the $z$-axis and infinity,
then they vanish identically. Assuming this has been shown, it follows that
the matrix fields must be equal for both solutions
$\Phi_i = \tilde \Phi_i$ for $i=1,2$. This may then be used to
prove that $\tilde g_{ab} = g_{ab}$ and $\tilde F_{ab} = F_{ab}$
as follows. First, the equality of the matrix fields
immediately implies $\tilde \chi = \chi, \tilde \alpha = \alpha, \tilde u = u, \tilde w = w$.
If $B = e^{u-2w} g_{ab} t^b \psi_2^a$, then we have $B \to 0$
at infinity and
\begin{equation}
\hat D_a B =2 re^{-4w} \, \hat \epsilon_a{}^b
\, \hat D_b \chi \, ,
\end{equation}
and similarly for the tilda solution. Thus, we have $\tilde B = B$. Finally, the norm of the
time-like Killing field $N = g_{ab} t^a t^b$ (and similarly for the tilda solution) satisfies
\begin{equation}\label{Ndef}
N= e^{-u+2w} B^2 - e^{-u-2w} r^2 \, ,
\end{equation}
from which it follows that $\tilde N = N$. Since $\psi_1^a$ is orthogonal to the other two Killing fields by assumption,
we also have $g_{ab} \psi_1^a \psi_2^b = 0 = g_{ab} t^a \psi_1^b$, and likewise for the tilda solution.
Hence, the inner products between all Killing fields are equal for both solutions.
Finally, it follows from the
equations eq.~\eqref{nudef} that also $\tilde \nu = \nu$, and it follows from
$F_{ab} \psi_1^a =  \nabla_a \alpha$ and our assumptions about the Maxwell field
that $\tilde F_{ab} = F_{ab}$. Altogether, this implies that the two solutions
coincide, as we desired to show. In fact, the metric and Maxwell field may locally
be written as
\begin{eqnarray}
ds^2 &=& - e^{-u- 2w}r^2 dt^2 + e^{-u + 2w}\left(d\phi_2 + Bdt\right)^2  + e^{-u +2w + 2\nu}\left(dr^2 + dz^2 \right) + e^{2u}d\phi^2_1 \nonumber \\
F &=&  \, d\alpha \wedge d\phi_1
\end{eqnarray}
in local coordinates such that $t^a = (\partial/\partial t)^a, \psi_i^a = (\partial/\partial \phi_i)^a$.

\medskip
\noindent
Thus, what remains is to be shown is that $\sigma_i$ is bounded. It is at this stage that we
must use our assumption that the interval structures and asymptotic charges of both solutions agree.
We must consider the behavior of $\sigma_i: \mr^3 \setminus \{z-{\rm axis}\} \to \mathbb H$ on (a)
near infinity (b) on the horizon, and (c) on the $z$-axis for both $i=1,2$.
We will consider these cases separately.

\medskip
\noindent
(a) In order to show that $\sigma_i$ are bounded near infinity, one uses that both metrics $\tilde g_{ab}$
and $g_{ab}$ are asymptotically flat near infinity (in $M$), with the same asymptotic charges
$\tilde m = m, \tilde J_1 = J_1 = 0, J_2 = \tilde J_2$, and the same electric charges $\tilde Q_{\rm E} = Q_{\rm E} = 0$. This can be used to show boundedness
of $\sigma_i$ near infinity in $\hat M$.

\medskip
\noindent
(b) On the open segment corresponding to the horizon, neither $e^w$ nor $e^u$ vanish, since both
Killing fields $\psi_i^a$ are non-vanishing by Prop.~2. Thus, $\sigma_i, i=1,2$ are bounded on the boundary segment of
$\partial \hat M$ corresponding to the horizon.

\medskip
\noindent
(c) On the boundary segments corresponding to a rotation axis, we must be most careful. We distinguish
boundary segments $(z_i, z_{i+1})$ where $\psi_1^a = 0, \psi_2^a \neq 0$ [corresponding
to the vector ${\bf v}_i = (1,0)$], boundary segments where $\psi_1^a \neq 0, \psi_2^a = 0$ [corresponding
to the vector ${\bf v}_i = (0,1)$], and corners where $\psi_1^a = 0 = \psi^a_2$.

Near points of the axis where
$\psi_1^a = 0, \psi_2^a \neq 0$,
we have $e^{2u} \to 0$ and $e^{2w} \to 0$ with $e^{2w-u}$ finite and non-zero, as the latter is
the norm of $\psi_2^a$ (and likewise for the tilda quantity). We first focus on this case.
 We immediately see that we have a potential problem in proving the boundedness of
 $\sigma_1$, see eq.~\eqref{sigdef1}, since the second term has $e^{u} e^{\tilde u}$ in the denominator, with
no compensating factors in the numerator as in the first term.
Clearly, $\sigma_1$ can only be finite if and only if $(\alpha - \tilde \alpha)^2$ goes to zero
near such points at least at the same rate as $e^{u} e^{\tilde u}$. Similarly, we also have a potential
problem in proving the boundedness of $\sigma_2$
see eq.~\eqref{sigdef2}, since the second term has $e^{2w} e^{2\tilde w}$ in the denominator, with
no compensating factors in the numerator as in the first term. Again,
$\sigma_2$ can only be finite if and only if $(\chi - \tilde \chi)^2$ goes to zero
near such points at least at the same rate as $e^{2w} e^{2 \tilde w}$.

We first determine the rate at which $e^u$ and $e^w$ tend to zero near the points where $\psi_1^a = 0, \psi_2^a \neq 0$. Since $e^{2w-u}$ is finite and non-zero near such points, it follows that $B$ is finite, too. From
the finiteness of $N$ and eq.~\eqref{Ndef}, it then also follows that $e^u = O(r)$, and
therefore that $e^{2w} = O(r)$.
Thus, in order for $\sigma_1$ and
$\sigma_2$ to be finite near such points, we must have $\tilde \alpha = \alpha + O(r)$ and $\tilde \chi = \chi
+ O(r)$.
We now prove that this is the case using the equality between the magnetic charges
$\tilde Q_{\rm M} = Q_{\rm M}$ and the angular momentum $\tilde J_2 = J_2$. For this, let $\zeta_1$ and
$\zeta_2$ be points on the boundary of the upper half plane $\hat M$ corresponding to points in the
manifold where $\psi_1^a = 0$. We can calculate the difference between $\alpha(\zeta_1)$ and
$\alpha(\zeta_2)$ by chosing an arbitrary path $\hat \gamma$ in the interior of the complex upper
half plane starting at $\zeta_1$ and ending at $\zeta_2$: Namely, since $f_a = \nabla_a \alpha$, we
have, in differential forms notation
\begin{equation}
\alpha(\zeta_1) - \alpha(\zeta_2) = \int_{\hat \gamma} \hat f \, .
\end{equation}
Now, it is possible to lift $\hat \gamma: [0,1] \to \hat M$ to a path $\gamma: [0,1] \to M$, i.e.,
$\hat \gamma = \pi \circ \gamma$, where $\pi$ is the projection from $M$ to the quotient space $\hat M$.
Let $C$ be the 2-surface in $M$ that is obtained by acting on points in the image of $\gamma$ with
the isometries generated by $\psi_1^a$, i.e.,
\begin{equation}\label{Cequal}
C := \{ (e^{2 \pi i t}, 0) \cdot \gamma(s) : \, \, s,t \in [0,1] \} \, .
\end{equation}
The images of the points $\gamma(0)$ and $\gamma(1)$ under the action
of this 1-parameter group isomorphic to $U(1)$ are again points, because $\psi_1^a |_{\gamma(0)} = 0 = \psi_1^a
|_{\gamma(1)}$. The image of any other point $\gamma(t), 0<t<1$ is a circle. Thus, it follows that the
2-surface $C$ is topologically a 2-sphere. If we now pick a local coordinate system near $C$ such that
$\psi_1^a = (\partial/\partial \phi_1)^a$, then we may write
\begin{equation}
\alpha(\zeta_1) - \alpha(\zeta_2) = \int_{\gamma} f = \frac{1}{2\pi} \int_C f \wedge d\phi_1 \, .
\end{equation}
where $\pi^* \hat f= f$, and where we have used in the second step that $\pounds_{\psi_1} f_a = 0$.
The term on the right side may now be manipulated using that $f_a = F_{ab} \psi_1^b$, showing that
\begin{equation}
\alpha(\zeta_1) - \alpha(\zeta_2) = \frac{1}{2\pi} \int_C F = \frac{1}{2\pi} Q_{\rm M}[C] \, .
\end{equation}
We may of course repeat the same argument for the tilda solution. Because the magnetic charges are the same
for the two solutions, it follows that $\alpha(\zeta) = \tilde \alpha(\zeta)$ up to a constant independent
of $\zeta$, for each $\zeta$ corresponding to a point where $\psi_1^a$ vanishes. Since that constant vanishes
at infinity by asymptotic flatness, it follows that $\sigma_1$ is finite near such points.

We would next like to show that the same statement holds true for $\sigma_2$.
This will follow if we can show that $\tilde \chi(\zeta) = \chi(\zeta) + O(r)$ for any $\zeta \in \partial \hat M$
not on the horizon segment. To show this, we first note that the twist 1-form $\omega$
vanishes on any axis, i.e. any point of $\partial \hat M$ not corresponding to the
horizon, by Prop.~1. Let $\zeta_1, \zeta_2 \in \partial \hat M$, and not
on the horizon segment, and take $\hat \gamma$ to be the curve
$\hat \gamma(t) = (1-t) \zeta_1 + t \zeta_2$ in $\hat M$. Then we have
\begin{equation}\label{chidef}
\chi(\zeta_1) - \chi(\zeta_2) = \int_{\hat \gamma} \hat \omega \, ,
\end{equation}
where $\pi^* \hat \omega = \omega$. If $\zeta_1, \zeta_2$ are both to the same side of the horizon, then
the above expression vanishes, while if they are on different sides, we find, by the same type of argument as above that
\begin{equation}
\chi(\zeta_1) - \chi(\zeta_2) = \frac{1}{(2\pi)^2} \int_{\mathcal H} *(d\psi_2) \, ,
\end{equation}
where $\psi_2^a$ has been identified with a 1-form via $g_{ab}$ and where $\mathcal H$ is a horizon
cross section in $M$. We would like to show that the quantity on the
right side is proportional to the angular momentum $J_2$. For this,
we pick a spacelike 4-surface $\Sigma$ in spacetime with interior boundary
$\mathcal H$ and boundary $S_\infty^3$ at infinity. By Gauss' theorem,
we can then write the quantity on the right side as
\begin{equation}
\int_{\mathcal H} \nabla_{[a} \psi_{2 \, b]} \, dS^{ab}
= J_2 + \int_{C} \nabla^b \nabla_{[a} \psi_{2 \, b]} \, dS^{a} \, .
\end{equation}
The integrand on the right side may be evaluated
standard identities for Killing vectors, the Einstein-Maxwell equations, as well as our
assumptions 1) and 2). We have
\begin{eqnarray}
\nabla^b \nabla_{[a} \psi_{2 \, b]} &=& \frac{1}{2} R_{ab} \psi_2^b \nonumber \\
&=&  \frac{1}{4}\left(F_{ac}F_{b}{}^c- \frac{g_{ab}}{6}F_{cd}F^{cd} \right) \psi_2^b \nonumber \\
&=&  -\frac{1}{48} \psi_1^b \psi_{1 \, b}^{} \xi^c \xi_c \psi_{2 \, a} =: \lambda \psi_{2 \, a} \, .
\end{eqnarray}
We may choose $\Sigma$ to be a surface defined by $T = const.$, where
$T$ is a time function that is invariant under the axial Killing fields\footnote{Such a function can be obtained from an arbitrary
time function $\tilde T$ by averaging $\tilde T$ over the compact group $\mathcal K$ of axial symmetries.}, i.e. in particular $\psi_2^a \nabla_a T = 0$.
Choosing now
an integration 4-form on $\Sigma$ by $\epsilon_{abcde} =
5 \nabla_{[a}T \epsilon_{bcde]}$, and letting $dS$ be the
integration element on $\Sigma$ associated with this 4-form, we see that
$\int_\Sigma \lambda \psi_{2 \, a} \, dS^a = \int_\Sigma
\lambda \psi_2^a \nabla_a T \, dS = 0$, as desired.
Since by assumption
$\tilde J_2 = J_2$, we conclude that $\tilde \chi(\zeta) = \chi(\zeta)$ on
any rotation axis, i.e. any point of $\partial \hat M$ not in the horizon segment. Since
the twist potential $\hat \omega$ also vanishes on $\partial \hat M$ except for the horizon
segment, it then follows from eq.~\eqref{chidef} that in fact even $\chi - \tilde \chi = O(r^2)$ near any
boundary segment corresponding to a rotation axis. Thus,
in summary, we have now shown that $\sigma_i, i=1,2$ has a finite limit for any point
$\zeta$ boundary of $\hat M$ where $\psi_1^a=0, \psi_2^a \neq 0$.

We must now consider the second case, i.e.,
points where $\psi^a_2=0, \psi_1^a \neq 0$. For such points,
$e^{2w-u} \to 0$, but $e^{u}$ finite and non-zero, so $e^{2w} \to 0$.
From the fact that $N$ is finite and non-zero near such points and
eq.~\eqref{Ndef} it can furthermore be seen that, in fact, $e^{2w} = O(r^2)$.
Thus, only $\sigma_2$ is potentially unbounded near such
points. However, we have already shown that $\tilde \chi - \chi = O(r^2)$ near any
point in $\partial \hat M$ which is not on the horizon segment, so this cannot happen. Thus,
$\sigma_i, i=1,2$ are bounded in that case, too.

Finally, we must consider the corners. Here we may invoke a continuity argument to show
that $\sigma_i$ are bounded.
Thus, when viewed as functions on
$\mr^3$, the functions $\sigma_i$ are solutions to eq.~\eqref{subharm}
that are bounded on the entire space $\mr^3$, including the $z$-axis.
As we have argued above, this is enough in order to show that
the two black hole solutions are identical. \qed

\medskip
\noindent
{\bf Remark:} The proof shows that the non-trivial 2-cycles [i.e., basis elements of $H_2(M)$] in the exterior of the spacetime may be obtained as follows. We know that the real axis bounding $\hat M$ is divided into intervals, each labeled with an
integer 2-vector $\v_i = (1,0)$ or $\v_i = (0,1)$. The different possibilities are summarized in the above table.
Now consider all possible curves $\hat \gamma_p, p=1,2,\dots$ in $\hat M$ with the property that $\hat \gamma_p$ starts
on an interval labeled $(1,0)$, and ends on another interval labeled $(1,0)$, with no interval with label $(1,0)$ in
between. If we now lift $\hat \gamma_p$ to a curve $\gamma_p$ in $M$, and act with all isometries generated by
$\psi_1^a$ on the image of this curve,
then we generate a closed 2-surface $C_p$ in $M$ [see eq.~\eqref{Cequal}],
which is topologically a 2-sphere for all $p$.
We may repeat this by replacing $\hat \gamma_p, p=1,2, \dots$ with a set of curves each
starting on an interval labeled $(0,1)$, and ending on another interval labeled $(0,1)$, with no interval with label $(0,1)$ in between. If we again lift these curves to curves in $M$, and act with all isometries generated by
$\psi_2^a$, then we generate a set of topologically inequivalent closed 2-surfaces $\tilde C_q, q=1,2, \dots$ in $M$, each of which is topologically a 2-sphere. It may be seen that the set of 2-surfaces $\{ C_p, \tilde C_q \}$ forms a basis of $H_2(M)$, and also of $H_2(\Sigma)$, where the 4-manifold $\Sigma$ is a spatial slice
going from infinity to the horizon (so that topologically $M=\mr \times \Sigma$). In this 4-manifold, we can compute
intersection numbers as $C_p: \tilde C_q = \pm 1$ or $=0$, depending on whether the corresponding
curves in $\hat M$ intersect or not. The rank of $H_2(\Sigma)=H_2(M)$ in the cases (i) through (iv) in the above table, and the intersection matrix $I_{pq} = C_p: \tilde C_q$ is therefore easily computed. This gives invariants of
the 4-manifold $\Sigma$ and hence of the exterior $M$ of the black hole.

Only the magnetic charges $Q_{\rm M}[C_p]$ enter in the proof of the above theorem. The magnetic charges
$Q_{\rm M}[\tilde C_q]$ are not needed and in fact vanish, due to assumptions 1) and 2) at the beginning of this
section. Thus, for the simplest interval structure $(0,1),(0,0),(1,0)$, there are no non-trivial magnetic
charges, and the unique solution within the class studied here
is completely specified by $J_2, m$. In fact, this unique solution is
the Myers-Perry black hole~\cite{Myers}, with vanishing Maxwell field.

\vspace{1cm}

\noindent
{\bf Acknowledgements:}  S. Y. would like to
thank the Alexander von Humboldt Foundation for a stipend, and
the Institut f\" ur Theoretische Physik G\" ottingen for its
kind hospitality. He also acknowledges financial support from
the Bulgarian National Science Fund under Grants MUF04/05 (MU 408) and VUF-201/06.

\end{document}